\documentclass[preprint,aps,superscriptaddress]{revtex4}
\usepackage{graphicx}
\usepackage{amsmath}
\usepackage[usenames]{color}
\usepackage{amssymb}
\newcommand{\be}{\begin{equation}}
\newcommand{\ee}{\end{equation}}
\newcommand{\bea}{\begin{eqnarray}}
\newcommand{\eea}{\end{eqnarray}}
\newcommand{\pr}{\partial}
\newcommand{\nno}{\nonumber}
\newcommand{\bse}{\begin{subequations}}
\newcommand{\ese}{\end{subequations}}

\begin{document}
\title{ Modified natural inflation: A small single field model with a large tensor to scalar ratio}
\author{Debaprasad Maity \footnote{debu.imsc@gmail.com} and Pankaj Saha
\footnote{pankaj.saha@iitg.ernet.in}}
\affiliation{Department of Physics, 
Indian Institute of Technology, Guwahati, India}

\begin{abstract}
In this paper we explored in detail a phenomenological model of modified single field natural inflation 
in light of recent cosmological experiments, BICEP2. Our main goal is to 
construct an inflationary model which not only predicts the important cosmological quantities such as $(n_s, r)$ compatible with
experimental observation, but also is consistent with the low energy effective theory framework.  
Therefore, all the fundamental scale apart from $M_p$ and quantities of our interest 
should be within the sub-Planckian region. In order to achieve our goal we modify the usual single field natural inflationary model
by a specific form of higher derivative kinetic term called kinetic gravity braiding (KGB).
One of our guiding principles to construct such a model is the constant shift symmetry of the axion.
We have chosen the form of the KGB term in such a way that it predicts
the required value of $n_s\simeq 0.96$ and a large tensor to scalar ratio $r> 0.1$. 
Importantly for a wide range of parameter space our model has sub-Planckian axion 
decay constant $f$ and the scale of inflation $\Lambda$. 
However, the reheating after the end of inflation limits the value of $f$ so that
we are unable to get $f$ to be significantly lower than $M_p$. 
Furthermore, we find sub-Planckian field excursion for the axion field $\Delta \phi \simeq f$ 
for the sufficient number of e-folding ${\cal N} \gtrsim 50 $. We also discussed in detail about the 
natural preheating mechanism in our model based on the recently proposed Chern-Simon coupling. We found 
this gravity mediated preheating is very difficult to achieve in our model. With our general analytic argument, 
we also would like to emphasize that Chern-Simons mediated preheating is very unlikely to happen in any slow roll inflationary model.

\end{abstract}

\maketitle

\newpage
\section{Introduction}\label{intro}
Based on various resent cosmological observations in the recent past, we are gradually 
reaching to the point that inflation must have happened in the early universe for a very short
period of time. Inflation is a paradigm in theoretical cosmology which had been proposed long time back 
 \cite{guth,linde1,steinhardt} in order to explain some conceptual problems in the standard 
big-bang theory. By now probably we are close to the view that inflation has happened before the standard
big-bang. However, soon after the introduction of this novel idea, it has been realised that it is very difficult 
to construct a consistent model of inflation. The reason behind this is our lack of understanding 
the nature beyond certain high energy scale. The very first question that is being asked
is the origin of inflation field itself. Simplest way is to introduce a simple scalar field called
inflaton. But this is not the end of the story. In the canonical formulation the potential 
which is needed to drive the exponential inflation has to be sufficiently flat. However quantum mechanically
it is very difficult to achieve such a flat potential. In addition to that from the observational point of view 
it has also been proved that the theory has to be in the super-Planckian regime where the model under consideration itself is questionable. 
Therefore, the theory and the experiments are proved to be very difficult to make compatible with each other. 
In this paper we will approach towards the goal to solve the aforementioned apparent tension between the theory and experiment.  
Keeping in mind the origin of the inflation field, axion field is supposed to be one of the most
natural candidates. In the context of particle physics, which is the most successful 
theory of nature, this axion field is assumed to play a key role in solving the strong
CP problem. In order to solve CP problem an extra $U(1)$ symmetry known as Pecci-Quinn (PQ) has been introduced.
The axion is produced after this symmetry is broken as a goldstone boson. It is this
symmetry also which makes the aixon field mass less before certain scale which is known as
QCD phase transition scale. After this QCD phase transition, the axion field develops a
very flat potential which is protected by the above mentioned symmetry from quantum mechanical 
correction. As we have mentioned above, the flat potential is one
of our main requirements to have sufficient amount of inflation. Therefore, keeping in mind the fact
that we do not have viable alternative to solve strong CP problem, standard model of particle
physics is assumed to be a natural framework to study inflation which is known as
natural inflation in the literature \cite{freese}. After the proposal of natural inflation
lots of work have been done in this direction to construct a viable model
of inflation. Even though it is one of the best theoretically motivated model, 
it is plagued by the super-Planckian effect. If we take the observational fact such as scale invariant power spectrum 
of the cosmological perturbation parametrised by so called spectral index $n_s$ measured quite 
accurately by PLANCK \cite{PLANCK} i.e.
\bea
n_s = 0.9603 \pm 0.0073 ,\nno
\eea
into account, the axion decay constant becomes $f \geq 3 M_p$. The very fact of having
super-Planckian scale related to any physical processes such as this axion decay constant
apart from the Planck scale, is incompatible with the low energy effective field theory framework. 
More precisely along with the Einstein general relativity (GR), quantum field theory itself
is ill understood at any scale greater than $M_p$ \cite{banks}. 
Therefore, there exists a clear tension between the cosmological observation and the inflationary model.
Furthermore, above the Planck scale the axionic shift symmetry itself may not be a good symmetry to solve the strong CP problem, or even if
it exits the dynamics of the axion field will severely get influenced by the unknown ultraviolet physics\cite{linde}. 
Hence to make axion inflation viable, significant amount of effort have been extended towards
understanding the possible mechanism to obtain sub-Planckian decay constant \cite{yamaguchi, hans}.
Following that lots of work have been done in understanding those models from more fundamental theoretical
point of view. One important earlier attempt was to formulate string inspired multi-field axion inflationary models \cite{kachru} motivated by
phenomenologically constructed assisted inflation \cite{assist}. Important to mention the other works in the similar line 
as ours were \cite{Hertzberg:2014sza, Ido:2014hi,sayantan}. In addition to this it is also worth mentioning a new kind of super-Planckian model called
monodromy inflation \cite{monodromy} inspired by string theory. This model also has got much interest
in the recent past. In this model constant shift symmetry is mildly broken by the linear potential due to 
non-perturbative effect in string theory. Another work which have some connection with ours \cite{germini}
is known as UV-protected inflation. 

So far most of the above discussions were based upon the CMB measurements namely Planck \cite{PLANCK}. 
And most importantly all the known single field slow roll inflationary models predict very small tensor power spectrum
in the effective field theory regime.
This was also consistent with the measurement of Planck experiment so far. 
But to everybody's surprise recently BICEP2 \cite{BICEP2} announced the detection of the rather large primordial B-mode
fluctuation in the cosmic microwave background (CMB). From which the large magnitude of tensor power spectrum is
expected. This quantity is quantified by the so called tensor-to-scalar ratio $r$ and BICEP2 tells us 
\bea
r = 0.20^{+0.07}_{-0.05} .\nno
\eea

Therefore, there exists an obvious tension between the Planck and BICEP2 results regarding the tensor power
spectrum which needs to have separate study. We will not discuss about this in the present paper. 
This particular result has also been questioned in the subsequent papers \cite{seljak} by 
taking into account the dust polarization map. This dust polarization effect might reduce
the above constraint to a significantly lower value which could become consistent with the Planck result.
Hence our analysis will not be confined with the BICEP2 observation. We will do our analysis with a 
wider point of view and predict the value of $r$ within the effective field theory framework.  
   The general argument given by D. Lyth \cite{Lyth} states that in the canonical formulation any model of single scalar field
inflation predicting tensor to scalar ratio $r>0.1$ should have super-Planckian field excursion.
Hence if we at least restrict ourselves to $ 0.2 > r > 0.1$, it is evident that 
constructing a model of inflation which predicts both the observed 
values of $(n_s, r)$ is very difficult in the framework of low energy effective theory. 
Clearly the physics beyond $M_p$ should be important which in turn ruin the predictive power of the original model
we started with. In this regard, monodromy inflation could be an interesting way out. In a recent interesting paper
the reference \cite{sami} showed that Lyth bound can be evaded in the quintessential inflationary model. 

In the present paper we will provide alternative scenario to resolve above mentioned super-Planckian problem. We will 
modify the usual canonical axion Lagrangian with a higher derivative term in the effective field theory
regime. By this modification we will show that all the important scales present in the model and 
the field excursion during inflation will be sub-Planckian. This higher derivative term of the axion field in the Lagrangian 
is known as kinetic gravity breading (KGB). 
Based on our previous work, we will study in detail a class of viable model of axion inflation in the light Planck and BICEP2. 
The particular term in the Lagrangian has already been used in the context of 
inflation \cite{yokoyama} and dark energy \cite{deffayet,felice}.
Following our previous work \cite{debuaxion,debubicep2}, we searched for 
a particular form of the higher derivative term which predicts the correct 
value of the important cosmological parameters $n_s$ and the value of $r$ within the detectable range
in the single field axion inflationary model. 
Moreover we do not have any super-Planckian problem as opposed to the usual model of single field
inflation. In summary we show that our model is consistent with the sub-Planckian axion decay constant $f < M_p$ and 
the scale of inflation $\Lambda < M_p$. 
The lower limit of the axion decay constant is constrained by 
requirement of the coherent oscillation of the axion field at the end of inflation. As we know this coherent oscillation will be 
essential for the reheating after the end of the inflation. With this sub-Planckian axion decay constant we computed the values of 
$(n_s,r)$ which are perfectly compatible with the observations by Planck and BICEP2. 
Furthermore our model also produces sufficient number of e-folding ${\cal N}$ within the range of
sub-Planckian evaluation of the axion field namely $\Delta \phi \simeq f$ during inflation.

At this stage we would like to emphasize that our important results of having all the scales 
below Planck seems to suggest that the model under consideration is in the effective field theory regime. 
Now the essential question that will automatically arise is the ultra-violet completion of our model. 
Answering this question is beyond our current objective of the paper. 
In addition, we also choose a particular class of model with some parameters which may get 
renormalized by loop effect unlike purely galileon model, where
the galileon symmetry, $\phi' = \phi + c + b_{\mu} x^{\mu}$, protects the underlying parameters from being renormalized.
Hence it would also be important to look into how our model parameters are corrected by ultraviolet modes
and effect our predictions.       
It is also important to note that having sub-Planckian field excursion should be taken
seriously in the effective field theory framework. Particularly as has been argued in \cite{Lyth}
it would be important if we want to associate the inflaton field as 
a visible sector field as an extension of the standard model, namely, minimal supersymmetric extension or
any other extension such as introducing neutrino masses, Peccei-Quinn(PQ) symmetry or even supergravity. All the 
above extensions are based on the effective quantum field theory where the field amplitudes 
are assumed to be very small at least compared to
the Planck scale. Our model is based upon the extension where axion is originating from
the breaking of PQ symmetry. Therefore, it is very important to have vacuum expectation
value of the inflaton field to be below Planck such that we have sub-Planckian field excursion.
A large class of well known inflationary models of inflation are based on extension of the standard model 
which are constructed in the effective field theory framework. Hence, having sub-Planckian field 
excursion for any inflationary model should be taken seriously unless we fully understand the fundamental theory
in the ultraviolet regime. String inspired models are supposed to be in that direction where 
super-Planckian or in other word high scale inflation has been consistently constructed.

 After we successfully reproduce the cosmological quantities of our interest compatible with the observations, we have 
studied in detail a gravity mediated pre-heating mechanism which has recently been proposed in \cite{CS-pre}. Interestingly,
the mechanism is very natural in the framework of axion inflation, where the axion is coupled
with the higher derivative gravitational Chern-Simons term. Even though the usual reheating scenario will work in our model 
once we chose the appropriate values of the coupling of the matter fields
with the axion, but it natural to have gravity mediated Chern-Simons preheating in the context any natural inflation. Furthermore, 
as has been argued, in this mechanism the minimal coupling of the matter field with the gravity will be sufficient to preheat the universe.
Therefore, we do not need to introduce any arbitrary coupling parameter in this scenario apart from the Chern-Simons coupling.
However, as we go along we found it very difficult to have successful preheating with this mechanism.

We organized our paper as follows, in section-II, we briefly introduce the model and corresponding 
cosmological quantities of our interests. We also discuss about the modification of the usual Lyth
bound due to the higher derivative term in our modified axion inflation scenario. We will see 
how the higher derivative kinetic term in the Lagrangian suppressed the change of the axion 
field value $\Delta \phi$ during inflation and making it sub-Planckian. We will see this numerically
in the subsequent section. In section-III, we will choose a specific class of braiding function $M(\phi)$.
The main motivation to choose those specific class of functions is to achieve our goal which is twofold. 
In one hand we will show numerically how our theory predictions of all the cosmological quantities $(n_s, r)$ 
are within the expected range coming from the cosmological observations such as Planck and BICEP2. 
On the other hand in the subsequent section-IV, we will see how the subsequent reheating phase after the inflation
will constrain our model parameters within the sub-Planckian range, such that our model is consistent
with the low energy effective field theory. Thereafter, we will discussed in detail the
natural preheating mechanism based on gravitational Chern-Simons coupling which has recently been proposed.
At the end we will summarize our result and discuss about future directions to work on.

\section{The model}
We will follow exactly our previous model where in addition to the usual canonical term we
also have higher derivative term called KGB for the axion field.
\bea
{\cal L} ~=~ \frac {M_p^2}{2}  R - X -  M(\phi) X \Box \phi
- \Lambda^4 \left(1 -\cos \left ( \frac {\phi}{f}\right)\right)
\label{action}
\eea
where
$
X = \frac 1 2 \pr_{\mu} \phi \pr^{\mu} \phi$ and $\Box = \frac 1 {\sqrt{-g}}\pr_{\mu}(
{\sqrt{-g}}\partial^{\mu})$, and  $\{f,\Lambda\}$ are the axion decay constant and  
axionic shift symmetry breaking scale respectively. $\Lambda$ is also known to be associated with the scale of inflation.
The very first discussion of this kind of higher derivative term known as galileon has been appeared in the context of long wave length 
modified gravity model \cite{rattazzi}. After this a spate of research work have been done in various aspects of 
its cosmological application. One of the most interesting properties of these kind of higher derivative terms is that
it does not lead to any ghost degrees of freedom as opposed to the usual higher derivative term. 

As has been mentioned in the introduction,
the guiding principle of constructing the basic form of the Lagrangian is  
galileon shift symmetry, $\phi' = \phi + c + b_{\mu} x^{\mu}$. This is the generalization of the axionic shift symmetry \cite{rattazzi}
which is originated from certain class of higher dimensional gravity model called DGP model.
This particular symmetry restricts the from of the Lagrangian to a class of specific higher derivative interaction term with 
constant coefficient such as above mentioned truncated Lagrangian eq.(\ref{action}) up to four derivative term, 
where the braiding function $M$ is constant, and more importantly this symmetry guarantees to have only two time derivative in the 
equation of motion avoiding the ghost instability. Because of this generalized shift symmetry all the model parameters such as constant $M$ 
in the present context has been shown to be radiatively stable \cite{porrati}. In this paper we consider a 
specific way of directly breaking this symmetry down to the constant shift symmetry by lifting the 
constant parameters $M$ to a function of the field obeying the constant shift symmetry. 
This particular way of breaking the symmetry does not lead to the ghost. But quantum loop effect
may renormalize the parameters which we will study in our future publication.   

One of 
our motivations to use this kind of higher derivative term is to cure the usual problem in axion inflation. And we will
see this is indeed the case.

With the usual background ansatz for the spacetime 
\be
ds^2 = -dt^2 + a(t)^2 (dx^2 +dy^2 + dz^2),
\ee
one gets the following Einstein's equations for the scale factor $a$ 
\bea
H^2 = - H \dot{\phi}^3 M(\phi)-\frac X 3 + \frac 2 3 X^2 M'(\phi)  + \frac {\Lambda^4}3
 \left(1 -\cos \left ( \frac {\phi}{f}\right)\right) 
\eea
and for the axion field 
\bea
\frac 1 {a^3} \frac d {dt} 
\left[a^3\left(1 - {3 H} {M} \dot{\phi} - 2 M' X \right)\dot{\phi}\right]
& =&  \pr^{\mu} \phi \pr_{\mu}(M' X) \nno\\ 
&-& 
\frac {\Lambda^4}{f} \sin \left ( \frac {\phi}{f}\right) .
\eea
Where, $H = {\dot{a}}/a$ is the Hubble constant.

Most of the computations of this section are followed from \cite{yokoyama}. The strategy is 
to identify the slow roll parameters as we usually do in the slow roll inflationary model. 
But there exists a clear difference between those two models. In the modified axion inflation i.e.
in the present case we have two different ways to achieve inflation. And this will in turn help
us to achieve our goal as will be clear from the subsequent discussion.   
Following \cite{yokoyama} and using the slow roll parameters which will be defined 
later, the scalar field equation takes the following form  
\bea \label{aeq}
3 H \dot{\phi} \left(1 - 3 M(\phi) H \dot{\phi} \right)+ 
\frac {\Lambda^4}{f} \sin \left ( \frac {\phi}{f}\right) =0 .
\eea
Here we clearly see that inflation can be driven by two different ways. Since usual axion inflation does not
solve the problems we mentioned. Our obvious choice would be the inflation driven by the KGB term. Hence,
the obvious condition that has to be satisfied for this is $|M(\phi) H \dot{\phi}|\gg 1$ leading
to the following inequality.
\bea
\tau = M(\phi) V'(\phi) = \frac {M(\phi) \Lambda^4} f \sin \left ( \frac {\phi}{f}\right) \gg 1 .
\eea  
If we want to have the potential driven inflation, we need to have very small value for the following
slow roll parameters, 
\bea \label{slowroll}
\epsilon &=& \frac {M_p^2} {2 f^2 \sqrt{\tau}} \frac {\sin \left( \frac {\phi}{f}\right)^2}
{\left(1- \cos \left(\frac {\phi}{f}\right) \right)^2} ~;~
\eta = \frac {M_p^2} {2 f^2 \sqrt{\tau}} \frac {\cos \left( \frac {\phi}{f}\right)}
{\left(1- \cos \left(\frac {\phi}{f}\right) \right)}  \nno\\
\alpha &=& \frac{ M_{p}}{2} \frac {M'}{M} \left (\frac {4 \epsilon^2}{\tau}\right)^{\frac
 1 4} ~~~;~~~\beta = \frac {M_{p}^2}{36} \frac {M''}{M} 
\left (\frac {4 \epsilon^2}{\tau}\right)^{\frac  1 2}  
\eea
For our later convenience we call the $M(\phi)$ as the KGB function. This potential
driven KGB inflation has been studied with a specific form the $M(\phi)$ in \cite{ohashi}.
Here our main goal is to identify the correct form of the KGB function such that our model
is perfectly viable from the theoretical as well as experimental point of view.

So far we have talked about the background evolution of the cosmological spacetime  
due to the KGB inflation and its various conditions. But most of the important cosmological 
observables such as large scale structure formation, CMB are originated from the tiny ripple on those 
above mentioned background solutions for the axion field and the background metric.
Those fluctuations are of quantum origin. Soon after the end of inflation those 
tiny fluctuations produces all the structure that we see in our observable universe 
through gravitational instability. The very first existence of those primordial fluctuation
have been observed in the CMB as a temperature fluctuation which is $\delta T \simeq 10^{-5}~~  {^o}K$. 
The quantity by which the temperature fluctuation is measured, is known as CMB power spectrum 
$P_{\cal R}$ where ${\cal R}$ is curvature perturbation, and an associated 
quantity called spectral index $n_s$. An another important component corresponding to those 
fluctuations is called gravitational waves. The amplitude of the
gravitational wave was theoretically as well as experimentally proved to be very small.
But to everybody's surprise recently it has been detected through the measurement of
B-mode polarization of the CMB at long angular scale by BICEP2 \cite{BICEP2}. 
From this particular measurement they derived another important cosmological parameter called 
scalar to tensor ratio $r$. So it became increasingly difficult to construct the model
of inflation which is consistent with the low energy effective theory in the usual canonical formalism.
The modification needs to be done towards a goal which we have discussed at the beginning. We will
see how the specific higher derivative term which is consistent with the low energy effective theory solves
the problem that we have just mentioned. 
The expressions for all those aforementioned cosmological quantities are \cite{yokoyama}
\bea
P_{\cal R} =  \frac {3 \sqrt{6}}{64 \pi^2}\frac {H^2}{M_p^2 \epsilon}
~~~;~~~ n_s = 1- 6 \epsilon + 3 \eta + \frac {\alpha} 2 ~~~;~~~n_T &=& - 2 \epsilon ~~;~~r =  - \frac {32 \sqrt{6}}{9} n_T .
\eea

As we have already mentioned in the introduction, the inflation was 
introduced to solve the homogeneity and flatness problem of the usual
Big-Bang model. Hence, in order to obtain present size of the observable universe we need to have 
sufficient amount of inflation. The amount of inflation is quantified by the so called e-folding number 
\bea
{\cal N}= \int_{t_1}^{t_2} H dt .
\eea
The current cosmological observations says that the value of ${\cal N} \gtrsim 50$. 
So this particular lower limit on ${\cal N}$ provides further constrains on the model parameters.
In terms of $M(\phi)$, the general expression for ${\cal N}$ turns out to be
\bea
{\cal N} &=&{\cal A} \int^{x_2}_{x_1} \frac{(1-\cos x) \sqrt{s^3 M(x)}}{\sqrt{\sin x}} dx,
\eea
where $s^3 M(x)$ is dimensionless KGB function. $s$ is KGB scale which will be defined later, and $ x = {\phi}/{f}$.
We have introduced a new constant ${\cal A} = \sqrt{\tau_0} (f/M_p)^2$.
We will see how this new combined parameter plays a very important role in solving our
problems in consistent with the observation. Important point to note here is that the
value of ${\cal A}$ also plays the main role in controlling the change of magnitude 
of the axion field during inflation. The upper limit
for the above integral $x_2 = \phi_2/f$ is fixed by setting the slow roll parameter
$\epsilon = 1$ and consequently the value of the lower limit $x_1$ of the above integral
can be obtained by taking a particular value of e-folding number consistent with the observation. 
At this point let me remind the readers again that in order to achieve $r>0.11$, the usual single field model of 
inflation is plagued by the super-Planckian field excursion of the inflation field known as Lyth bound.
In the following sub-section we will see how the Lyth bound is modified due to the higher derivative term
and consequently that results into the sub-Planckian field excursion of the axion field during the full period
of inflation.

\subsection{Modified Lyth bound} 

As we have discussed above the e-folding number can be expressed as
 \bea
{\cal N}= \int_{t_1}^{t_2} H dt =\int_{t_1}^{t_2} \frac {H}{\dot{\phi}} d\phi = \frac 1 {M_p} 
\int_{\phi_{end}}^{\phi_{in}} \frac {\tau^{\frac 1 4}} {\sqrt{2 \epsilon}} d\phi.
\eea 
Where we have used the slow roll condition eq.(\ref{slowroll}). One can check that the important
modification in the expression for e-folding number is $\tau(\phi)$ function as opposed to the 
usual case in derivation of the Lyth bound. We will see that this is the term which save us from the 
super-Planckian field excursion of the axion field during inflation. So the bound on the field will come from
following relation  
\bea
{\cal N} \lesssim \frac {\Delta \phi} {M_p} 
\left|\frac {\tau(\phi)^{\frac 1 4}} {\sqrt{2 \epsilon}}\right|_{max} = \frac {\Delta \phi} {M_p} 
\left|\frac {\tau_{max}^{\frac 1 4}} {\sqrt{2 \epsilon_{min}}}\right| .
\eea
Where we have define the prefix $"max"$ to be the value of the functions under consideration
at the on-set of inflation. $\Delta \phi = |\phi_{in}-\phi_{end}|$. $\phi_{in}$ and $\phi_{end}$ are
the value of the axion field at the beginning and at the end of the inflation. From the above expression
it is obvious that the minimum value of the slow roll parameter
$\epsilon_{min}=\epsilon(\phi_{in})$ will contribute to the modified Lyth bound. If we
use the value of $r$ at the beginning of the inflation in the above expression, the bound on $\Delta \phi$ 
turns out to be
\bea
{\Delta \phi} \gtrsim ~({\cal N} {M_p}) \left|\frac {\sqrt{2 \epsilon_{min}}}  {\tau_{max}^{\frac 1 4}}\right| = 
\frac f {\sqrt{{\cal A}}} \frac {{\cal N}}{{\cal T}_{max}} \sqrt{\frac { 9 r} { 36 \sqrt{6}}},
\eea
where, 
\bea
{\cal T}_{max}= (s^3 M(x_{in}) \sin x_{in})^{\frac 1 4}. \nno
\eea
Hence, one can clearly see that the bound on the change of axion field value during inflation is suppressed by the 
same combined parameter ${\cal A}$. More interestingly we see that by suitably choosing the 
value of ${\cal A}$ one can make all the result consistent with the observation and still get the 
sub-Planckian $\Delta \phi$. From our following detail numerical analysis we can clearly see that.
In the next section we will discuss in detail a specific class of models and the corresponding 
cosmologically relevant quantities such as $(n_s, r, \Lambda)$ and how
all those are depending on the model parameters $\{f,s,\Delta \phi\}$ for specific  
e-folding number ${\cal N} = 50, 60$. The discussion for higher value of ${\cal N}$ will be straightforward. 
We will study in detail the relevant parameters and quantity $\{\phi_1/f, \phi_2/f, n_s, r, \Delta \phi \}$.

\subsection{Specific model: Form of $M(\phi)$}

In the model of axion inflation, one of our first goal is to reduce the value of $f$
by some mechanism. One way to achieve this is to dynamically reduce its value by introducing another scale such as ours KGB scale s.
For example in the reference \cite{hans} it has been done by 
introducing multiple axion fields with their respective sub-Planckian axion decay constants and the dynamics 
of the combined system enhance the effective decay constant to super-Planckian value which leads to the scale invariant power spectrum. 
In the model with the higher derivative term like our KGB modified axion inflation, 
we will see that it is the KGB function $M(\phi)$ and its associated specific higher derivative 
term which as an alternative mechanism predicts all the cosmologically relevant quantities compatible 
with the Planck and BICEP2 measurements even with the sub-Planckian $f$. Furthermore the same mechanism 
predicts the sub-Planckian field excursion for the axion field during inflation. Our main guiding principle 
of choosing the form of $M(\phi)$ is to keep the constant shift symmetry
intact. Hence we will be restricted to some form of $M(\phi)$ which contains 
$\sin\left(\phi/f\right)$ and $\cos\left(\phi/f\right)$ function.
 
 We will consider the following particular class of KGB function of form
\bea \label{kgbM}
M(\phi) = \frac 1  {s^3} {\sin^p x} \left[1-\cos x \sin^2 x \right]^q.
\eea
where $p$ is odd integer and $q$ is any integer . 
In this paper we will discuss in detail three possible choices of $p= \{5,7,9\}$ and show how those function will
lead us to the required results. We have also checked higher values of $p$ and quite obviously found the
same qualitative behaviour. But importantly it is further lowering down the limiting value of the
axion decay constant $f$. We will see that only for $p>5$, $f$ becomes sub-Planckian consistent with
the reheating. For $p=5$, even though we get $f$ little higher than $M_p$ but $\Delta \phi$ during inflation is
still sub-Planckian. For every value of $p$, we choose some value of $q$ and show how
the value of $\{n_s,r\}$ depend on $q$. Therefore, we will consider three classes of models
based on the value of $p$. We call $s$ as KGB scale. We will see how this single KGB scale $s$ provides us all
the aforementioned interesting results for $p>5$. The functional form that we have considered came out from many trial 
functions and indeed a fairly complicated choice. Interestingly enough, we will see that this choice gives us robust result
in terms experimental observation as well as theoretical point of view. Hence it would be 
interesting to understand the origin of this function in terms of more fundamental theory. 
The main properties that appear to be very important of this class of function is its behaviour near
the minimum of the potential and specific form inside the bracket. The function $M(\phi)$ should go to zero very rapidly
near the minimum compare to the potential such that after the end of inflation, the dynamics 
will be governed by the usual canonical kinetic term and the potential for sub-Planckian decay constant. 
Expanding the above function $M(\phi)$ and the inflationary potential $V(\phi)$ near $\phi =0$,
\bea
M(\phi) \approx \frac {1}{s^3} \left(\frac {\phi}{f}\right)^p ~~~;~~~  
V(\phi) \approx \Lambda^4 \left(\frac {\phi}{f}\right)
\eea

\begin{figure}[t!]
\includegraphics[width=3.400in,height=2.00in]{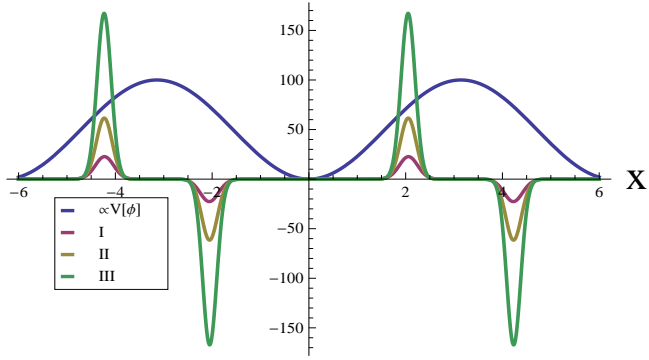}
\caption{\label{fig1} Qualitative plot for the potential $V(\phi)$ and $M(\phi)$ for three
different models up to a constant factor related to the amplitude.}
\end{figure}
\begin{figure}[t!]
\includegraphics[width=5.400in,height=1.20in]{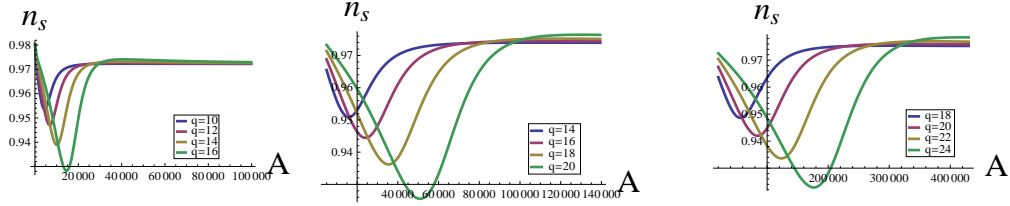}
\caption{\label{na} Behaviour of the spectral index $n_s$ with respect to the derived parameter 
${\cal A}$ for three different functional form of $M(\phi)$. In this plot we have taken $p= 5, 7, 9$
from left to right respectively and e-folding number ${\cal N} = 50$. We will have the similar 
behaviour for ${\cal N} = 60$.
occurs in the region I}
\end{figure}
\begin{figure}[t!] 
\includegraphics[width=5.400in,height=1.20in]{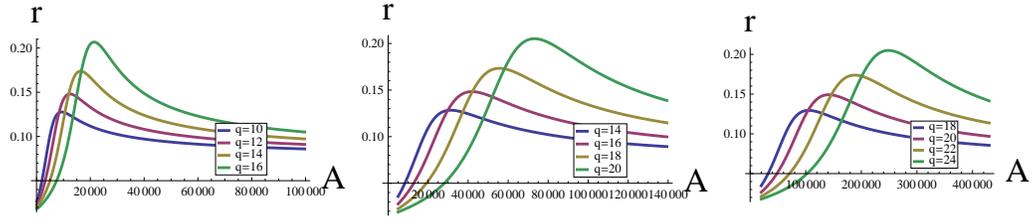}
\caption{\label{ra} Behaviour of the scalar to tensor ratio $r$ with respect to the derived parameter 
${\cal A}$ for three different functional form of $M(\phi)$. In this plot we have taken $p= 5, 7, 9$
from left to right respectively and e-folding number ${\cal N} = 50$. We will have the similar 
behaviour for ${\cal N} = 60$}
\end{figure}
\begin{figure}[t!] 
\includegraphics[width=5.400in,height=1.20in]{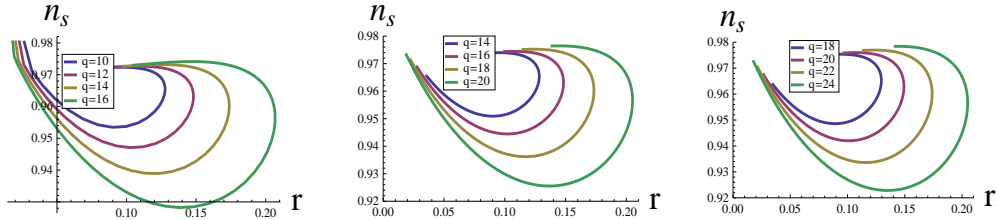}
\caption{\label{rn} Behaviour of the spectral index $n_s$ with respect to the scalar to tensor ratio $r$ 
for three different functional form of $M(\phi)$. In this plot we have taken $p= 5, 7, 9$
from left to right respectively and e-folding number ${\cal N} = 50$. We will have the similar 
behaviour for ${\cal N} = 60$}
\end{figure}


we show that for $p>3$ only we have coherent oscillation for sub-Planckian axion decay constant.  
We have also seen that the odd property of the function plays an important role. Also because
of this property inflation can happen on both side of the potential. In the next section we will
study three different class of functions in detail. Keeping in mind the experimental values
of the cosmological parameters $n_s$ and the possible value of $r$ within the expected range, 
we plotted $({\cal A} ~vs~ n_s)$ in fig.\ref{na}, 
$({\cal A}~ vs ~r)$ in fig.\ref{ra} and $({r ~vs~ n_s})$ in fig.\ref{rn} for each model.
The behaviour of particular plot for three different models are qualitatively same.
From those plots one of the interesting points that we noticed that for our specific choice of the KGB 
function, near around the experimental values of $n_s =0.96$, scalar to tensor ratio $r$ always 
takes value around its maximum depending upon the value of $q$.

{\bf Model-I}: In this model we choose the form of the braiding functions to be 
\bea \label{kgbf}
M(\phi) = \frac 1  {s^3} {\sin^5 x} \left[1-\cos x \sin^2 x \right]^{q}.
\eea
We particularly choose three different values of $q = 12, 14, 16$ and see how the values of $(n_s,r)$
depend on those q-values. For ${\cal N} = 50$ , we see from the fig.\ref{na} that for each value of $q$, we have
two different values of ${\cal A}$ such that $n_s = 0.96$. For all value of $q$, spectral
index converges to a constant value $n_s \simeq0.972$ and scalar to tensor ration converges to 
$r\simeq 0.091$ for large value of ${\cal A}$. Most interestingly in the $(r ~vs~ n_s)$ plot, if we fix the value of the spectral
index to be the observed central value $n_s\simeq 0.96$, one gets 
\[ \begin{array}{c c c}
p=5 & {\cal N}=50 & {\cal N}= 60 \\
\begin{array}{c}
\hline
q  \\
\hline
10\\
12 \\
14\\
16 \\
\hline
\end{array}
&
\begin{array}{c|c|c|c|c}
\hline
 {\cal A} & r & x_1 & x_2 & \frac {\Lambda}{M_p} \\
\hline
 7300 &  0.124 & 0.89 & 0.202 & 0.010\\
 11500 & 0.147 & 0.84 & 0.185 & 0.011\\
 16300 & 0.174 & 0.82 & 0.172 & 0.012\\
 22300 & 0.206 & 0.80 & 0.162 & 0.013\\
\hline
\end{array}
&
\begin{array}{c|c|c|c|c}
\hline
{\cal A} & r & x_1 & x_2 & \frac {\Lambda}{M_p}\\
\hline
5300  & 0.077 & 1.053 & 0.022 & 0.0086\\
10900 & 0.112 & 0.931 & 0.187 & 0.00997\\
16900 & 0.140  & 0.884 & 0.171 & 0.0105\\
124700 & 0.172 & 0.842 & 0.158 & 0.0116\\
\hline
\end{array}
\end{array}\]

{\bf Model-II}: In this model we choose the following braiding functions  
\bea \label{kgbf}
M(\phi) = \frac 1  {s^3} {\sin^7 x} \left[1-\cos x \sin^2 x \right]^{q}.
\eea
As one can see that in model-II, the braiding function goes to zero near 
the minimum of the potential much more rapidly than that of the previous one. Hence, naturally we need 
to consider higher value of $q$ in order to met the experimental values of $(n_s,r)$
mentioned in the introduction. We consider $q = 16, 18, 20$. The behaviour of 
all the various plots are qualitatively same. For this model let us summarize the main results in the 
following table   

\[ \begin{array}{c c c}
p=7 & {\cal N}=50 & {\cal N}= 60 \\
\begin{array}{c}
\hline
q  \\
\hline
14 \\
16  \\
18\\
20 \\
\hline
\end{array}
&
\begin{array}{c|c|c|c|c}
\hline
 {\cal A} & r & x_1 & x_2 & \frac {\Lambda}{M_p} \\
\hline
26000 & 0.125 & 0.868 & 0.219 & 0.011\\
39400 & 0.148 & 0.835 & 0.204 & 0.011\\
56000 & 0.173 & 0.814 & 0.192 & 0.012 \\
76000 & 0.204 & 0.803 & 0.183 & 0.012\\
\hline
\end{array}
&
\begin{array}{c|c|c|c|c}
\hline
{\cal A} & r & x_1 & x_2 & \frac {\Lambda}{M_p}\\
\hline
22600 & 0.087 & 0.975 & 0.225 & 0.009\\
40000 & 0.116 & 0.899 & 0.203 & 0.010\\
60000 & 0.142 & 0.864 & 0.190 &  0.011 \\
85000 & 0.171 & 0.838 & 0.179 &  0.012\\ 
\hline
\end{array}
\end{array}
\]    
where again we have considered the value of $n_s = 0.96$

{\bf Model-III}: For this case similarly we choose the form of the braiding functions to be 
\bea \label{kgbf}
M(\phi) = \frac 1  {s^3} {\sin^9 x} \left[1-\cos x \sin^2 x \right]^{q}.
\eea
We consider the values of $q = 20, 22, 24$. Considering the value of $n_s = 0.96$ again we summarize our result 
in the following table. 

\[ \begin{array}{c c c}
p=9 & {\cal N}=50 & {\cal N}= 60 \\
\begin{array}{c}
\hline
q  \\
\hline
18 \\
20 \\
22\\
24 \\
\hline
\end{array}
&
\begin{array}{c|c|c|c|c}
\hline
 {\cal A} & r & x_1 & x_2 & \frac {\Lambda}{M_p} \\
\hline
92000  & 0.127 & 0.851 & 0.232 & 0.010\\
135000 & 0.149 & 0.829 & 0.219 & 0.011 \\
189000 & 0.174 & 0.814 & 0.209 & 0.012 \\
256000 & 0.204 & 0.835 & 0.200 & 0.012 \\
\hline
\end{array}
&
\begin{array}{c|c|c|c|c}
\hline
{\cal A} & r & x_1 & x_2 & \frac {\Lambda}{M_p}\\
\hline
88000  &  0.095  & 0.927 & 0.234 & 0.0096 \\
140000 & 0.118   & 0.884 & 0.218 & 0.010\\
206000 & 0.142 & 0.857 & 0.20617 & 0.011\\
289000 & 0.170 & 0.837 & 0.196 & 0.012\\
\hline
\end{array}
\end{array}
\]

From the above tables and figures for three different models we see the general trend of 
the value of $r$ with the increasing value of $q$. Hence clearly if we increase the value of $q$ further, 
the value of $r$ will increase. All these values of $r$ can be made compatible with the observation of BICEP2 
considering various effects on the B-mode 
polarization of CMB coming from the background dust, cosmological magnetic field etc. As we have discussed earlier
and clearly see from the above tables for all the three models that the change in axion field value during the inflation 
turns out to be 
\bea
\Delta \phi  = (x_1 - x_2) \times f  <  f .
\eea
Therefore if we have sub-Planckian axion decay constant $f$ or very close to unity in Planck unit, 
the axion will have sub-Planckian field excursion with the required value of e-folding number. 
Consequently our model will not have any serious trans-Planckian problem. 
Once we have determined all the above parameters,
by using the expression for the scalar power spectrum 
\bea \label{pr}
P_{\cal R} =  \frac {{\cal A}\sqrt{6 }}{32 \pi^2} 
\left( \frac {\Lambda}{M}\right)^4 
\frac{(1-\cos x_1)^3 \sqrt{ s^3 M(x_1)}}{\sin ^{\frac{3}{2}} x_1} \simeq 2.4\times 10^{-9},\nno
\eea
we get $\Lambda \approx 0.012 $ in Planck unit for all the above models.

At this point we would like to mention and also obvious from the 
similar behaviour of the aforementioned three different models that one can keep increasing 
the value of $p = 11, 13, \cdots$ and get perfectly
viable model of inflation. Since qualitative behaviour of the results are same provided
the suitable choices of $q$, we will not discuss those any further. 
However we will see in the subsequent section that as we increase the value of $p$, one can reduce 
the lower limit of the axion decay constant $f$. So far we have discussed on how one can
met the observational results and obtain the sub-Planckian $\Delta \phi$ by choosing the value of our combined parameter ${\cal A}$.
However one should remember that in order to fix the value of $(f,s)$  
we used the following relation 

\bea \label{ratio}
\frac {s^3}{f^3} = \frac 1 {{\cal A}^2} \frac {\Lambda^4}{M_p^4}.
\eea

Unfortunately this equation provides us an infinite number of possibilities to have the sub-Planckian value of $(f, s)$ which satisfy 
the above equation. So one can easily get all the constants of our model of interest to be sub-Planckian.
What we have not discussed so far is the behaviour of the axion after the end of inflation.
This is one of our motivations to choose the aforementioned specific class of KGB functions $M(\phi)$ Eq.(\ref{kgbf}) 
such that we get coherent oscillation of the axion field after the end of the inflation. 
As we know the coherent oscillation of the inflaton field is essential to reheat our universe after the 
end of inflation. In the subsequent section we will discuss about how this requirement of coherent oscillation at the end of 
inflation constraints our model parameters. We also discuss about the evolution of the energy density of the axion field 
and the most natural pre-heating scenario in our model which has recently been discussed in \cite{CS-pre}. 

\section{Reheating}

In this section we will first discuss about the dynamics of the axion field after the end of inflation and show that
condition for coherent oscillation will provide us further constrain on the region of parameter space of $(f,s)$.
As we know in the usual reheating scenario, due to the coherent oscillation of the inflaton field, all the standard model
matter field directly or indirectly coupled with the inflaton, will be produced through the parametric instability after the inflation. 
In order to check whether the inflaton field coherently oscillates after the end of the inflation, we have numerically 
solved the Einstein's equations with the slow roll boundary conditions 
that has been derived before. We have numerically checked that the lower bound of $f$ depends on the value of ${\cal A}$
for a particular choice of $(p,q,\Lambda)$. In the following tables we provide the lower bound on $f$ above which we
have coherent oscillation of the axion field after the end of inflation.

\[ \begin{array}{ccc}
 \begin{array}{c}
p=5 , {\cal N}=50\\
\begin{array}{c|c|c}
\hline
q & {\cal A} & f\gtrsim  \\
\hline
10 & 7300  & 1.15\\
12 & 11500 & 1.26 \\ 
14 & 16300 & 1.35\\
16 & 22300 & 1.45\\
\hline
\end{array}
\end{array}&
\begin{array}{c}
p=7 , {\cal N}=50\\
\begin{array}{c|c|c}
\hline
q & {\cal A} & f \gtrsim \\
\hline
14 & 26000 & 0.84\\
16 & 39400 & 0.90 \\ 
18 & 56000 & 0.95\\
20 & 76000 & 1.00\\
\hline
\end{array}
\end{array}&
\begin{array}{c}
p=9 , {\cal N}=50\\
\begin{array}{c|c|c}
\hline
q & {\cal A} & f \gtrsim \\
\hline
18 & 92000  & 0.68\\
20 & 135000 & 0.71 \\ 
22 & 189000 & 0.74\\
24 & 256000 & 0.77\\
\hline
\end{array}
\end{array}
\end{array}
\]

\begin{figure}
\includegraphics[width=6.00in,height=1.50in]{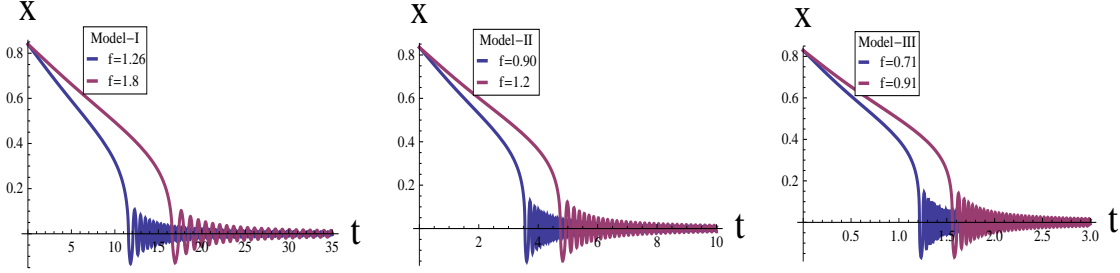}
\caption{\label{reheat} It shows the coherent oscillation of the axion field after the end 
of the inflation for all the three models. We have plotted for ${\cal N} =50$. One
can clearly see that the axion speed is increasing as go from Model-I to Model-III (from left panel to right panel). 
We use $q= 12, 16, 20$ and correspondingly ${\cal A} = 11500, 39400, 135000$ for Model-$\{I, II, III\}$ respectively. 
Time $t$ is measured in unit of KGB scale $s$.}
\end{figure}

We have computed the value of $f$ in units of Planck. Therefore,
the reheating after the end of inflation sets a lower limit on the value of 
the axion decay constant depending on our specific choice of KGB function. 
We see that for $p=5$ even though the value of $f$ is little above the Planck value, the
axion field excursion $\Delta \phi$ is still in the sub-Planckian regime. 
For the fixed value of spectral index, as we increase the value of
${\cal N}$, lower bound on $f$ slightly increases. We find it very difficult to reduce the value of $f$ 
significantly lower with the class of KGB functions we have studied. 
For instance if we further increase the value of $p = 11, 13$, the lower bound on the axion decay constant 
turn out to be $f\approx 0.6, 0.5$ respectively.
Moreover we clearly see from the fig.(\ref{reheat}) that once we 
fix the parameters $(p,q,\Lambda)$, the speed of the axion field during the 
inflation depends on the valued of $f$. As we increase the value
$f$, the inflationary time period also increases. Hence in general the KGB driven
inflation is not really a slow roll inflation once we restrict our $f$ to be in the 
sub-Planckian regime as compared to the usual super-Planckian natural inflation. 
We can clearly see from the fig.(\ref{reheat}) that the inflationary time period is significantly 
reduced from Model-I to Model-I. Hence from the behaviour we have studied so far it will be 
very difficult to further reducing the lower bound of the axion decay constant which causes 
increasing the speed of the axion during inflation. It would be interesting to construct a model
where the axion speed decreases with increasing value of $f$.   

\begin{figure}
\includegraphics[width=6.00in,height=1.50in]{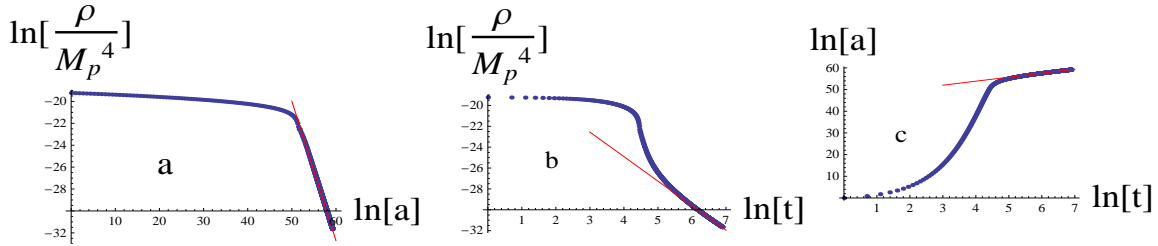}
\caption{\label{rho} In these plots we show how the energy density $\rho_{\phi}$, and the scalar factor $a(t)$ 
behave with respect to the time (two panels on the right ($b,c$). The left panel $a$ shows the behaviour of $\rho_{\phi}$ with respect to
$ln[a]$. All the red line are fitting curve in the asymptotic limit. All these behaviours are independent of the models
under consideration.}
\end{figure}

As we have seen, after the end of the inflation the axion field coherently oscillates even with the sub-Planckian 
value of $f$. During the evolution we also numerically solved for the effective energy density of the
axion field, 
\bea
\rho_{\phi} =  - 3 H \dot{\phi}^3 M(\phi)- X  + 2  X^2 M'(\phi)  + {\Lambda^4}
 \left(1 -\cos \left ( \frac {\phi}{f}\right)\right) 
\eea
as shown in the fig.(\ref{rho}). The fall off behaviour of the effective energy density $\rho_{\phi}$ and the scale factor $a(t)$ 
after the end of the inflation are found to be the same as expected for all the models under consideration. 
By fitting those fall of behaviour (red lines) we numerically found 
\bea
\rho_{\phi} \propto \frac 1 {a^{1.27}}~~~;~~~~\rho_{\phi} \propto {t^{1.62}}~~~;~~~~a \propto t^2 .
\eea

Hence the energy density of the axion field reduces very slowly with the cosmological expansion as compared to the
usual radiation and matter energy density. This behaviour also is in sharp contrast with the usual slow roll
inflation scenario where the time average energy density of the inflaton under consideration would behave like a matter field
after the end of inflation, mainly because of the quadratic nature of the potential near its minimum. So it would be
interesting to look at if there exists any distinguishable feature of our model on the simultaneous 
reheating mechanism as opposed to the usual single field inflationary model. In the context our model we will
be discussing about a new kind of preheating mechanism which has recently been proposed \cite{CS-pre}. As we
are studying modified axion inflation, it is natural to consider the coupling of axion field with the gravitational Chern-Simons term.
It is known that this kind of parity violating coupling leads to the enhancement of the amplitude of the gravitational
perturbation. Using this mechanism, the authors have proposed a natural pre-heating scenario where all the 
matter field is produced without any unnatural coupling with the inflaton. 
We have discussed in detail their mechanism in the context of our model. We will try to analytically argue that it is actually very difficult to
to have successful preheating by just using that mechanism.  

\subsection{Gravitational Chern-Simons preheating}

As we have already mentioned above and also form the fundamental theoretical point of view
such as string theory, the most natural another higher derivative term that can play an important role in the context
of natural inflation is 
\bea
{\cal L}_{int} \simeq \frac {g} {f} \phi R {\tilde R}
\eea
where the pseudo scalar axion field couples to the gravitational Chern-Simons term
\bea
R {\tilde R} &=&  \tilde{R}^{\nu}{}_{\nu}{}^{\alpha \beta} \, R^{\nu}{}_{\mu \alpha \beta}\, ,  \hspace{1.5cm} 
\tilde{R}^{\nu}{}_{\nu}{}^{\alpha \beta} := \frac{1}{2}\, \epsilon^{\alpha \beta \gamma \delta} R^{\mu}{}_{\nu\gamma \delta}. \nonumber 
\eea
$\epsilon^{\mu\nu\alpha\beta}$ is the usual 4-dimensional Levi-Civita tensor. 
$g$ is the free coupling parameter. Important point to note that this interaction term does not 
contribute to the background evolution as $\tilde{R} R$ vanishes identically in the
homogeneous background. Hence the main contribution coming from this term will be in 
the analysis of perturbation. Since this gravitational Chern-Simons term violates parity, it is well known that
due to this interaction the left and right handed propagating degrees of freedom of the gravitational wave
will follow different dynamics. During inflation one of the modes gets exponentially suppressed and other one 
gets enhanced. It will be clear from our discussion below how this happening. 

To the linear order in cosmological perturbation theory, the Fourier transformed gravitational wave equation 
turns out to be
\bea \label{gwave}
\left(M_p^2 + \frac {k ~ g ~\eta_A} {f ~a} \dot{\phi} \right)\left(\ddot{h}^k_A + 3 H \dot{h}^k_A + \frac {k^2}{a^2} h^k_A \right) 
=- \frac {k~ g ~\eta_A}{f~ a} (\ddot{\phi} - H\dot{\phi})~\dot{h}^k_A,
\eea  
and the corresponding Fourier transformed equation for the matter field say $\theta$ which is minimally coupled with the gravity 
will take the following from 
\bea \label{thetaeq}
\ddot{\theta}_k + 3 H \dot{\theta}_k + \left(\frac {k^2}{a^2} + m^2 \right) \theta_k = - \sqrt{\frac {2}{\pi}} \frac {1}{a^2} 
\int \mbox{Re}\left[\sum_{A=R,L} {k}_A^{'2}~ h^k_A({\bf k}-{\bf k}')\right]~  \theta({\bf k}')~ d^3 {\bf k}' 
\eea 
In the above equations of motion, we have used the circular polarization basis \cite{circular} of the gravitation wave
which particularly decouples the gravitational wave equations. $\eta_A = \pm 1$ for right(R) and left(L) circularly 
polarized gravitational wave respectively. $k$ is the magnitude of momentum of the particle produced. $m$ is the mass of the
particle field $\theta$. ${k}_{R,L} = k_x \pm i k_y$. In the above eq.(\ref{gwave}) we have ignored the back-reaction of the 
matter field on the gravity wave. We are particularly interested in quantum mechanical production of particles. Therefore, 
we promote the above wave equations to the evolution equation of a particular mode function of the
quantum field $h$ and $\theta$ with a definite momentum $k$. We will numerically solve the above set of mode equations
in our inflaton background $\phi$ in the oscillatory regime and compute the evolution of the particle number density 
$n_k$ for the matter field $\theta$ with a definite momentum ${\bf k}$. 
The well known expression for the particle number density is \cite{kofman} 
\bea
n_k = \frac {\omega_k}{2} \left( \frac {|\dot{\theta}_k|^2} {\omega_k^2} + {|{\theta}_k|^2}\right) - \frac 1 2 ,
\eea
where as usual $\omega_k$ is the time dependent frequency of a particular mode ${\bf k}$, 
\bea
\omega_k =  \sqrt{\frac {k^2}{a^2} + m^2} = \sqrt{{k_{phy}^2} + m^2}
\eea
where $k_{phy}$ is the physical momentum of the particle.

As we see, one has an integro-differential eq.(\ref{thetaeq}) to solve where in the integration part we need to do 
a three dimensional momentum integral. In order to make our computation simple we fix the value of $k_x, k_y$ and do
the one dimensional integral along the $k_z$. Before we solve for the particle production for specific parameter values, 
let us try to understand the equation and its behaviour from the analytic point of view. As has been pointed out in \cite{CS-pre},
in order to have successful preheating one needs to consider large initial amplitude for the gravitational wave during the
oscillation period after the inflation. In the following subsection we want to discuss about this enhancement mechanism closely.   

\subsubsection{Amplitude enhancement mechanism during inflation}   

It is already well known \cite{circular} that because of the Chern-Simons coupling the right circular polarization mode of the 
gravity wave will get enhanced during the inflation. In order to understand this analytically let us write down the equation for the gravitational
wave eq.(\ref{gwave}) in a slightly different form as follows:
\bea \label{modgwave}
\left(M_p^2 + \frac {k ~ g ~\eta_A} {f ~a} \dot{\phi} \right)\left[\ddot{h}^k_A + 3 H\left( 1 - \frac {k~ g ~\eta_A}{M_p^2 f~ a} 
\frac{\dot{\phi}}{3 (1 + \frac {k ~ g ~\eta_A} {M_p^2 f ~a} \dot{\phi})} \right) \dot{h}^k_A + \frac {k^2}{a^2} h^k_A \right] 
=0
\eea

We want to understand the behaviour of the gravitational wave during the inflation. Hence we have used the slow roll condition $\ddot{\phi}/(H \dot{\phi}) \ll 1$
in the above equation. Usually people analyse the above equation in conformal time in which the equation becomes simpler.
However we will try to understand this in cosmic time in which it is easier to analyse the pure time dependent friction term of the above equation. 
The main controlling parameter for the enhancement of the mode is 
\bea
B = \frac {k ~ g ~} {M_p^2 f ~a} \dot{\phi}.
\eea
At this point we would like to point out \cite{instability} that the modes which are kinetically unstable because
of wrong sign kinetic term in the Lagrangian will have exponential enhancement due to the Chern-Simons term. This can be clearly
seen from the above equation as we will discuss next. During inflation $\dot{\phi} < 0$ or in other word $B < 0$ assuming Chern-Simons coupling is positive. 
Therefore kinetic instability happens only for $\eta_R = 1$ with the following condition at a particular instant of time
\bea
|B| > 1 \equiv \frac {k}{a} = k_{phy} >  \frac {M_p^2 f} {g |\dot{\phi}|} = k_{cs},
\eea 
where $k_{phy}$ is the physical wave number at a particular instant of cosmic time $t$. 
Now let us consider the time dependent friction term from the above eq.(\ref{modgwave}), we call it $F$
\bea
F =   1 - \frac {\eta_A~B}{3 (1 + \eta_A ~ B)} 
\eea   
The real enhancement of the gravity wave mode will occur only when $F<0$. As one can clearly see, for $\eta_L = -1$, $1\geq F \geq 2/3 $. Therefore,
for left circularly polarized gravity wave, Chern-Simons term will play as a source of anti-friction. Hence
it helps to reduce the suppression of the amplitude of a particular mode under consideration caused by the usual Hubble parameter.
This is where in the literature, it is said that the mode amplitude of the left handed gravity wave is exponentially enhanced by the
Chern-Simons term. But strictly speaking at the end of the inflation the amplitude of the mode under consideration can not 
grow greater that unity due to the Hubble friction term.

On the other hand for $\eta_R = 1$, we have two different situations for
\bea
F = 1 + \frac {\eta_A~|B|}{3 (1 + \eta_A ~ |B|)} .
\eea   
If $|B| < 1$, the Chern-Simons term will further enhance the suppression factor for the amplitude of the right circular polarization mode 
during inflation. On the other hand when $|B| > 1$ which is the condition for the kinetic instability one can easily compute that
F becomes negative only for $ 1 < B < 3/2 \implies  1 < k_{phy}/k_{cs} < 3/2 $. Hence for $|B| > 3/2$, 
F again becomes positive. Therefore, what we can conclude from the above discussion 
is that the amplitude of the left circular polarization mode will get pure enhancement during the inflation only within a very 
small physical momentum window. Otherwise all the other modes will be suppressed by the Hubble expansion overcoming the anti-friction 
effect coming from the Chern-Simons term. Therefore we see that a very small fraction of the modes within the kinetically
unstable modes will have exponential enhancement during inflation. Now the question we would like to ask is whether 
the amplitude of all the modes within that specific window $ 1 < k_{phy}/k_{cs} < 3/2 $ will have exponential enhancement or not.
As is quite clear that during inflation we have exponential expansion of the spacetime. Hence even though we started with a specific
physical moment $k_{phy}$ within that window, during inflation the enhancement only happens within a very very short period of time.
Otherwise it will have suppression due to the effective positive friction term. What we have discussed so far have been checked
numerically. We also want to emphasize that as per as our above analytic arguments and numerical checks are concern, our conclusion
is actually model independent in the slow roll inflationary regime.  

Hence to summarize, for the left circularly polarized gravity wave $h_L$, the Chern-Simons term will exponentially enhance the amplitude
of the mode under consideration but can never overcomes the exponential suppression caused by the Hubble parameter $H$. Therefore, effectively
we still have large amplitude suppression and not possible to get very large amplitude at the end of the inflation.
On the other hand for the right circular polarized mode $h_R$ we have a very narrow range of the momentum modes which will get exponentially amplified
due to Chern-Simons term overcoming the Hubble suppression. Hence amplitude larger than unity at the end of the inflation could be 
achieved. But unfortunately all those modes are kinetically unstable. Hence vacuum will be unstable which has been discussed in the reference \cite{instability}.
It has been emphasized in the reference \cite{CS-pre} that in order to have successful preheating one needs to have large amplitude of 
the mode under consideration as an initial condition. In the following discussion we will first verify their results in our cosmological background in the 
oscillatory regime and discuss about some important points. Then we try to argue that it is very difficult to get such a large amplitude 
of the gravitational wave at the end of the inflation even with the Chern-Simons coupling. 

\subsubsection{Numerical solution and particle production}  

\begin{figure}
\includegraphics[width=6.00in,height=1.50in]{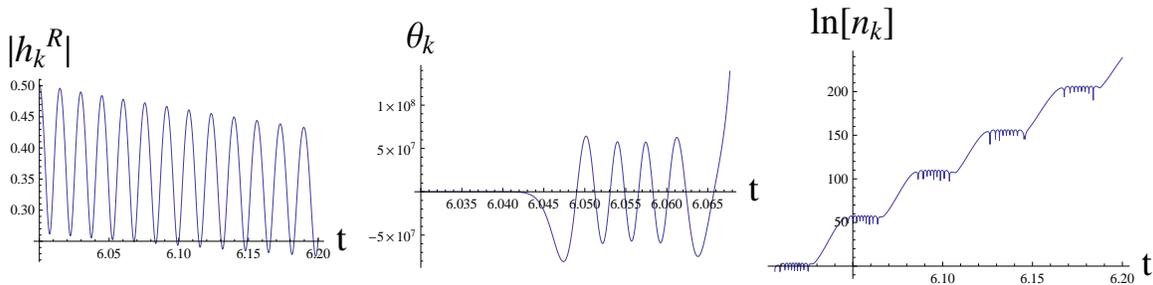}
\caption{\label{particle} Behaviour of gravitational wave $|h_k$, the matter field $\theta_k$ and particle number 
density $n_k$ in time t which is measured in unit of s. We have chosen $k_x = 0, k_y = 170, k_z=130$ in unit of s. With this choice
of momentum the Chern-Simons contribution is almost negligible. Hence setting $g=0$ does not influence these results.}
\end{figure}       
 
 In order to check our numerical calculation we first compute the quantum mechanically produced particle for
certain mode $n_k$ during the oscillatory regime with the following set of parameter choices. We reproduced 
the result of the reference \cite{CS-pre} for the following particular choices.
We choose the model-II and $q= 14, {\cal A} = 26000$ such that ${\cal N} = 50, n_s=0.96, r = 0.124$
and $f = 0.84, s = 2.25 \times 10^{-6}$. We set $t=0$ as the beginning of the inflation, then from numerical calculation 
we found the inflation ends at around $t \sim 4 $ cosmic time in unit of $s$. We have done our numerical calculation
for the preheating staring at $t = 6$ by normalizing the scale factor $a(6) = 1$. In the fig.\ref{particle}, we have
plotted the particle density for a particular value of momentum $k_{phy} = k/a(6)$. We want to emphasize that 
for the above particle production Chern-Simons coupling has no significant effect as for this particular choice of momentum Chern-Simons contribution
is very small. Hence we also get the same plot setting $g=0$. Only quantity which matters in this analysis is the
initial amplitude of the gravitational wave and its momentum. We have checked this with a large range of momentum and its initial 
amplitude. We also check that if the initial amplitude is very low, there is no particle production. 

Therefore, the important requirement that we need in order to have the parametric resonance in the matter field evolution
is a reasonably large amplitude of the gravitational wave as an initial condition during the oscillation period. 
After verifying this fact and their numerical analysis of \cite{CS-pre} we went for the full numerical solution.
We tried to numerically solve the full set of perturbation equations starting from the beginning of inflation. We use
usual Bunch-Davis vacuum as an initial condition at $t=0$.   
But unfortunately we were not able to get the numerical solution with large amplitude for those modes which will contribute to the particle
production during reheating. Let us provide an analytic argument behind this negative result for our model following the discussion in the previous
sub-section. As we have discussed in the previous sub-section we consider the mode which will have exponential enhancement
due to the Chern-Simons term and the corresponding effective friction term is as follows 
\bea 
F = 1 - \frac {|B|}{3 (1 + |B|)} .
\eea  
Hence, the maximum enhancement will happen to those mode for which $|B| \gg 1$. This condition leads to the following constraint 
on the physical momentum 
\bea \label{kcs}
k_{phy} \gg k_{cs} =  \frac {M_p^2 f} {g |\dot{\phi}|} =  \frac {M_p^2} {g s |\dot{X}|} \implies k \gg \frac {a(t) M_p^2} {g s |\dot{X}|}
\eea
where $\dot{X} = \dot{\phi}/(s f)$ is dimensionless quantity. Now as we know the initial amplitude of 
a particular mode is fixed by the usual Bunch-Davis vacuum as follows:
\bea
h_A = \lim_{t \rightarrow 0 }  \frac 1 {a(t) \sqrt{2 k}} e^{- i k \int \frac 1 {a(t)} dt} \left(1 + \frac i { k \int \frac 1 {a(t)} dt}\right),
\eea
 In the above equation we assume that deep inside the Hubble volume,
the scalar field remains almost constant. And just to have leading order correction we consider approximately
the de-Sitter vacuum. This does not effect much on our final conclusion. Therefore, the amplitude of the gravitational wave
depends on a particular mode under consideration. If we use the above eq.(\ref{kcs}), one can find the amplitude of the mode as
\bea
|h_A| = \sqrt{\frac {g~ s~ |\dot{X}|} {2 a(t) n^2 M_p^2} } \left | 1 + \frac i { k \int \frac 1 {a(t)} dt}\right |.
\eea
We are computing each parameter in Planck unit. We introduce a parameter $n$ which controls the value of $k$. For large enough value
of $n$ above condition eq.(\ref{kcs}) is satisfied. Our main interest is the boundary condition at the beginning of inflation namely at $t=0$ and 
ignoring the second part of the above equation, the amplitude of a particular mode comes out to be
\bea
|h_A|(t=0) = \sqrt{\frac {g ~s_p^2 |\dot{X}|} {2 n} }
\eea
where $a(0) =1, s_p = s/M_p$. In the above equation we measure the momentum in $s$ unit which is natural for our model.
In general our model predicts $s_p \approx 10^{-6}$. Since we consider slow roll inflation, and $\dot{X}$ is dimensionless, one also has $|\dot{X}| < {\cal O}(1)$.
In addition to that if we consider $g \sim 1$, the initial amplitude of the gravitational wave which can contribute to the particle production at the 
end of the inflation will be
\bea
|h_A|(t=0) \lessapprox \frac {10^{-6}}{n}. 
\eea
Hence, starting with this amplitude at the beginning of inflation, the background inflation essentially 
reduces the amplitude of that particular mode further down. Only contribution of the Chern-Simons term would be to
reduce this suppression. Hence it is unlikely to have preheating by using the Chern-Simons term at least in our model.

For the {\it Strong coupling case when $g \gg 1$} we were also unable to find particle production unless we choose large amplitude
at which we already found the particle production even setting $g=0$ like we have seen in the fig.\ref{particle}. All our numerical
computation makes us believe that the above conclusion is not only true for our case but probably true in any general slow roll inflation. 
It is very difficult to get large amplitude of a gravitational wave after the end of inflation.

\section{Conclusions}
 
In this paper we have tried to constructed a phenomenological model of modified natural inflation. 
The main modification of our model is a specific form of the higher derivative term for the axion
field in the usual axion Lagrangian. This kind of modification was first introduced as a
local modification of gravity in \cite{rattazzi}. After that lots of work have been done
in various aspects of cosmology. The main term which plays the essential role in our discussion is a function associated with 
the above mentioned higher derivative term called braiding function $M(\phi)$. In this paper our goal was twofold. 
In addition to satisfy the observational predictions coming from various cosmological 
experiments, we also wanted to construct the model which is consistent with the low energy 
effective theory. Keeping those motivation in mind, 
we phenomenologically choose the form of the $M(\phi)$ in such
a way that we can reproduce all the important results of inflationary cosmology
 and also it is consistent with the effective theory. The main principle which guided us to choose the special form
of those function is the constant shift symmetry of the original axion Lagrangian. 
 Soon after the introduction of single field inflationary model,
it has been realised through the observational perspective that such a model is indeed difficult to construct in the 
canonical formulation at least with the single field inflaton. Hence, we employ a specific form of non-canonical 
term in the Lagrangian for the axion field. This particular 
term is coined as kinetic gravity braiding in the literature. 
Thanks to its particular form which is not plagued with the ghost problem.
The result that we have got is robust in the perspective of low energy
effective field theory. The main result that we want to emphasize once more
is that if the axion inflation is driven by the higher derivative term of the Lagrangian,
for the observed central value of $n_s = 0.960$, we will have the following 
one particular choice of all the other parameters for $r \sim 0.147$ 
for ${\cal N} =50$. Interestingly all the quantities are sub-Planckian which was
one of our main motivations to achieve. We also want to emphasize that with this value of
parameters the axion field oscillates coherently after the end of the inflation,
which will lead to successful reheating of the universe. 
Hence, reheating after the end of inflation sets a lower limit on the value of
the axion decay constant $f$ depending on our specific choice of KGB function.

\[ \begin{array}{c|c|c|c|c|c}
 \hline
p & {\cal A} & \frac {f} {M_p}  & \frac {\Delta \phi}{M_p} & \frac {s}{M_p} & \frac {\Lambda} {M_p} \\
\hline
5 & 11500  &  1.26  &  0.825  &  6.20 \times 10^{-6}  &  0.011  \\
7 & 39400  &  0.90   &  0.568  &  1.96  \times 10^{-6}  &  0.011 \\ 
9 & 135000 &  0.71  &  0.433  &  6.84 \times 10^{-7}  &  0.011 \\
\hline
\end{array}
\]

Recently a new preheating mechanism has been proposed in \cite{CS-pre}, where 
gravitational wave plays the main role in reheating the universe after the inflation.
In this mechanism gravitational Chern-Simons term plays the main role. As has been pointed
out in that reference, the usual minimal coupling of any matter field with the
gravity is sufficient to reheat the universe by parametric resonance. According to 
their claim, due to Chern-Simons term, the amplitude of a particular chiral gravitational wave gets
exponentially enhanced due to exponential expansion of the background spacetime. This enhanced 
gravitational wave will lead the parametric resonance to the matter field and reheat the universe
after the inflation. In this paper we have studied their proposal in detail 
in our modified axion inflation scenario. Since the dynamics of the gravitational wave amplitude
is important in the aforementioned preheating mechanism, we have discussed explicitly 
how the amplitude of the gravity wave is modified separately by the usual Hubble friction and 
the effective friction/anti-friction coming from the gravitational Chern-Simons term. 
We have seen that effectively amplitude of the initial gravitational wave will not get 
enhanced after the end of the inflation in the regime where there is no ghost mode.
The Chern-Simons term will only help to reduce the Hubble friction effect. Now the question 
remains whether that Chern-Simons anti-friction effect will really help to reheat after the inflation.
For that we have first numerically checked the result of \cite{CS-pre} which has been shown in the fig.\ref{particle}.
But unfortunately we were unable to produce their result starting from beginning of the inflation.
So we concluded that the large amplitude of the gravitational wave at the end of inflation is not possible to achieve 
even with the Chern-Simons term. In order to understand this we have given an analytic explanation.
We know that the initial amplitude of the gravitational wave is fixed by the Bunch-Davis vacuum and deep
inside the Hubble horizon it behaves like $|h_k| \propto 1/\sqrt{k}$. Therefore with the increasing
the frequency the initial amplitude of that particular mode decreases. What we found is the modes which
acquire large modification due to the Chern-Simons term will start to evolve with the initial amplitude 
$|h_k| < 10^{-6}/n$. Where $n$ is real number which is much greater than unity. Therefore as it evolves the gravitational amplitude 
will get exponentially reduced during the inflation by the effective positive friction term 
\bea
F = 1 - \frac {|B|}{3 (1 + |B|)} \nno.
\eea
Hence, at the end of the inflation we found it very difficult to find the large amplitude  of the gravitational wave which can 
trigger the parametric resonance and reheat the universe. We also conjecture that our conclusion should be true for
any slow roll inflationary model. It would be interesting if one can introduce some mechanism which can 
produce large amplitude gravitational wave after the end of inflaton. Therefore, the usual reheating 
is still the best viable scenario to reheat the universe. It would be interesting to construct a model
where gravitational amplitude can get pure enhancement overcoming the Hubble friction. After the end of
inflation this new mechanism can provide us a large amplitude of the gravitational wave which can lead to the 
gravity mediated preheating even without the coherent oscillation of the inflaton field.

In summary our model is within the effective field theory regime and hence our results are robust.
However, because of limitation of our theoretical construction we were unable to get the axion decay constant to be 
significantly lower than $M_p$. As we have mentioned earlier it would be interesting to look into this problem and
construct a model which can circumvent this issue. We defer this for our future study. 
Furthermore, by our current analysis we were not able to fix all our model parameters completely. 
Therefore, some more cosmological observables such as non-gaussianity, CMB anomaly in the large scale 
could be interesting to look at, and which can help us fixing that free parameter.
In order to achieve our goal we have introduced some complicated higher derivative terms in the action. It would be
interesting to understand the origin of those functions from more fundamental theoretical point of view.

{\bf Acknowledgement}\\
We are very thankful to our HEP-group members to have vibrant academic discussions.

\end{document}